\documentclass[12pt]{article}
\begin{document}
\title{From Bosonic Strings to Fermions}
\author{B.G. Sidharth\\
International Institute for Applicable Mathematics \& Information Sciences\\
Hyderabad (India) \& Udine (Italy)\\
B.M. Birla Science Centre, Adarsh Nagar, Hyderabad - 500 063 (India)}
\date{}
\maketitle
\begin{abstract}
Early string theory described Bosonic particles at the real life Compton scale. Later developments to include Fermions initiated by Ramond and others have lead through Quantum Super Strings to M-theory operating at the as yet experimentally unattainable Planck scale. We describe an alternative route from Bosonic Strings to Fermions, by directly invoking a non commutative geometry, an approach which is closer to experiment.
\end{abstract}
\section{Bosonic Strings}
Let us begin with T. Regg's work of the 1950s \cite{reg,roma,tassie} in which he carried out a complexification of the angular momentum and analysed particle resonances. As is well known, the resonances could be fitted by a straight line plot in the $(J,M^2)$ plane, where $J$ denotes the angular momentum and $M$ the mass of the resonances. That is we have
\begin{equation}
J \propto M^2,\label{ex}
\end{equation}
Equation (\ref{ex}) suggested that not only did resonances have angular momentum, but they also resembled extended objects. This was contrary to the belief that elementary particles were point like. In fact at the turn of the twentieth century, Poincare, Lorentz, Abraham and others had toyed with the idea that the electron had a finite extension, but they had to abandon this approach, because of a conflict with Special Relativity. The problem is that if there is a finite extension for the electron then forces on different parts of the electron would exhibit a time lag, requiring the so called Poincare stresses for stability \cite{rohr,barut,feynman}.\\
In this context, it may be mentioned that in the early 1960s, Dirac came up with an imaginative picture of the electron, not so much as a point particle, but rather a tiny closed membrane or bubble. Further, the higher energy level oscillations of this membrane would represent the ``heavier electrons'' like muons \cite{dirac}.\\
Then, in 1968, G. Veneziano came up with a unified description of the Regge resonances (\ref{ex}) and other scattering processes. Veneziano considered the collision and scattering process as a black box and pointed out that there were in essence, two scattering channels, $s$ and $t$ channels. These, he argued gave a dual description of the same process \cite{ven,venezia}.\\
In an $s$ channel, particles A and B collide, form a resonance which quickly disintegrates into particles C and D. On the other hand in a $t$ channel scattering  particles A and B approach each other, and interact via the exchange of a particle $q$. The result of the interaction is that particles C and D emerge. If we now enclose the resonance and the exchange particle $q$ in an imaginary black box, it will be seen that the $s$ and $t$ channels describe the same input and the same output: They are essentially the same.\\
There is another interesting hint which we get from Quantum Chromo Dynamics. Let us come to the inter-quark potential \cite{lee,smolin}. There are two interesting features of this potential. The first is that of confinement, which is given by a potential term like
$$V (r) \approx \sigma r, \quad r \to \infty ,$$
where $\sigma$ is a constant. This describes the large distance behavior between two quarks. The confining potential ensures that quarks do not break out of their bound state, which means that effectively free quarks cannot be observed.\\
The second interesting feature is asymptotic freedom. This is realized by a Coulumbic potential
$$V_c (r) \approx - \frac{\propto (r)}{r} (\mbox{small}\, r)$$
$$\mbox{where}\, \propto (r) \sim \frac{1}{ln(1/\lambda^2r^2)}$$
The constant $\sigma$ is called the string tension, because there are string models which yield $V(r)$. This is because, at large distances the inter-quark field is string like with the energy content per unit length becoming constant. Use of the angular momentum - mass relation indicates that $\sigma \sim (400 MeV)^2$.\\ 
Such considerations lead to strings which are governed by the equation \cite{walter,fog,bgskluwer,st}
\begin{equation}
\rho \ddot {y} - T y'' = 0,\label{e1}
\end{equation}
\begin{equation}
\omega = \frac{\pi}{2l} \sqrt{\frac{T}{\rho}},\label{e2}
\end{equation}
\begin{equation}
T = \frac{mc^2}{l}; \quad \rho = \frac{m}{l},\label{e3}
\end{equation}
\begin{equation}
\sqrt{T/\rho} = c,\label{e4}
\end{equation}
$T$ being the tension of the string, $l$ its length and $\rho$ the line density and $\omega$ in (\ref{e2}) the frequency. The identification (\ref{e2}),(\ref{e3}) gives (\ref{e4}), where $c$ is the velocity of light, and (\ref{e1}) then goes over to the usual d'Alembertian or massless Klein-Gordon equation.\\
Further, if the above string is quantized canonically, we get
\begin{equation}
\langle \Delta x^2 \rangle \sim l^2.\label{e5}
\end{equation}
Thus the string can be considered as an infinite collection of harmonic oscillators \cite{fog}. Further we can see, using equations (\ref{e2}) and (\ref{e3}) and the fact that
$$\hbar \omega = mc^2$$
that the extension $l$ is of the order of the Compton wavelength in (\ref{e5}), a circumstance that was called one of the miracles of the string theory by Veneziano \cite{ven}.\\
It must be mentioned that the above considerations describe a ``Bosonic String'', in the sense that there is no room for the Quantum Mechanical spin. This can be achieved by giving a rotation to the relativistic quantized string as was done by Ramond \cite{uof,ramond}. In this case we recover (\ref{ex}) of the Regge trajectories. The particle is now an extended object, at the Compton scale, rotating with the velocity of light. Furthermore in superstring theory there is an additional term $a_0$, viz.,
\begin{equation}
J \leq (2\pi T)^{-1} M^2 + a_0 \hbar, \, \mbox{with}\, a_0 = +1 (+2) \, \mbox{for \, the \, open\, (closed)\, string.}\label{e6}
\end{equation}
The term $a_0$ in (\ref{e6}) comes from the Zero Point Energy. Usual gauge bosons are described by $a_0 = 1$ and gravitons by $a_0 = 2$.\\
It is also well known that string theory has always had to deal with extra dimensions which reduce to the usual four dimensions of physical spacetime when we invoke the Kaluza-Klein approach at the Planck Scale \cite{kaluza}.\\
All these considerations have been leading to more and more complex models, the latest version being the so called M-Theory. In this latest theory supersymmetry is broken so that the supersymmetric partner particles do not have the same mass as the known particle. Particles can now be described as soliton like branes, resembling the earlier Dirac membrane. M-Theory also gives an interface with Black Hole Physics.  Further these new masses must be much too heavy to be detected by current accelerators. The advantage of Supersymmetry is that a framework is now available for the unification of all the interactions including gravitation. It may be mentioned that under a SUSY transformation, the laws of physics are the same for all observers, which is the case in General Relativity (Gravitation) also. Under SUSY there can be a maximum of eleven dimensions, the extra dimensions being curled up as in Kaluza-Klein theories. In this case there can only be an integral number of waves around the circle, giving rise to particles with quantized energy. However for observers in the other four dimensions, it would be quantized charges, not energies. The unit of charge would depend on the radius of the circle, the Planck radius yielding the value $e$. This is the root of the unification of electromagnetism and gravitation in these theories.\\
In M-Theory, the position coordinates become matrices and this leads to, as we will see, a noncommutative geometry or fuzzy spacetime in which spacetime points are no longer well defined \cite{madore}
$$[x,y] \ne 0$$
From this point of view the mysterious $M$ in M-Theory could stand for Matrix, rather than Membrane.\\
So M-Theory is the new avatar of QSS. Nevertheless it is still far from being the last word. There are still any number of routes for compressing ten dimensions to our four dimensions. There is still no contact with experiment. It also appears that these theories lead to an unacceptably high cosmological constant and so on.\\   
The non-verifiability of the above considerations and the fact that the Planck scale $\sim 10^{20}GeV$ is also beyond forseeable attainment in collidors has lead to much criticism even though it is generally accepted that the ideas are promising.
\section{Fuzzy Spacetime and Fermions}
With the above background, we now attempt to use Bosonic Strings which are at the real world Compton scale to obtain a description of Fermions without going to the Planck scale. We saw above that Bosonic particles could be described as extended objects at the Compton scale. Given a minimum spacetime scale $a$, it was shown long ago by Snyder that,
$$[x,y] = (\imath a^2/\hbar )L_z, [t,x] = (\imath a^2/\hbar c)M_x, etc.$$
\begin{equation}
[x,p_x] = \imath \hbar [1 + (a/\hbar )^2 p^2_x];\label{e3z}
\end{equation}
It may be mentioned that (\ref{e3z}) is compatible with Special Relativity. Furthermore if $a^2$ in (\ref{e3z}) is neglected, then we get back the usual canonical commutation relations of Quantum Mechanics. This limit to an established theory is another attractive feature of (\ref{e3z}).\\
However if order of $a^2$ is retained then the first of equations (\ref{e3z}) characterize a completely different spacetime geometry, one in which the coordinates do not commute. This is a noncommutative geometry and indicates that spacetime within the scale defined by $a$ is ill defined, or is fuzzy \cite{madore}. As we started with a minimum extention at the Compton scale, let us take $a = (l, \tau)$.\\
Then the above conclusion is in fact true, because as discussed in detail \cite{cu,weinberg}, by virtue of the Heisenberg Uncertainty Principle, there are unphysical superluminal effects within this scale. In fact there is a non zero probability for a particle at $(\vec{r}_1, t_1)$ to be found at $(\vec{r}_2, t_2)$, whatever be $t_1$ and $t_2$, as long as
$$0 < (\vec{r}_1 - \vec{r}_2)^2 - (t_1 - t_2)^2 \leq l^2$$
Another way of seeing this is by starting from the usual Dirac coordinate \cite{pdirac}
\begin{equation}
x_\imath = \left(c^2 p_\imath H^{-1}t\right) + \frac{1}{2} c\hbar \left(\alpha_\imath - cp_\imath H^{-1}\right) H^{-1}\label{e9}
\end{equation}
where the $\alpha$'s are given by 
\begin{equation}
\vec \alpha = \left[\begin{array}{ll}
\vec \sigma \quad 0\\
0 \quad \vec \sigma
\end{array}
\right]\quad \quad ,\label{e10}
\end{equation}
the $\sigma$'s being the usual Pauli matrices. The first term on the right side of (\ref{e9}) is the usual Hermitian position coordinate. It is the second or imaginary term which contains $\vec{\alpha}$ that makes the Dirac coordinate non Hermitian. However we can easily verify from the commutation relations of $\vec{\alpha}$, using (\ref{e10}) that
\begin{equation}
[x_\imath , x_j] = \beta_{\imath j} \cdot l^2\label{eA}
\end{equation}
In fact (\ref{eA}) is just a form of the first of equations (\ref{e3z}) and brings out the fuzzyness of spacetime in intervals where order of $l^2$ is not neglected.\\
Dirac himself noticed this feature of his coordinate and argued \cite{pdirac} that our physical spacetime is actually one in which averages at the Compton scale are taken. Effectively he realized that point spacetime is not physical. Once such averages are taken, he pointed out that the rapidly oscillating second term in (\ref{e9}) or zitterbewegung gets eliminated.\\
We now obtain a rationale for the Dirac equation and spin from (\ref{eA}) \cite{bgscsf,bgsnc}. Under a time elapse transformation of the wave
function, (or, alternatively, as a small scale transformation),
\begin{equation}
| \psi' > = U(R)| \psi >\label{e8a}
\end{equation}
we get
\begin{equation}
\psi' (x_j) = [1 + \imath \epsilon (\imath x_j \frac{\partial}{\partial x_j}) + 0 (\epsilon^2)] \psi
(x_j)\label{e9a}
\end{equation}
Equation (\ref{e9a}) has been shown to lead to the Dirac equation
when $\epsilon$ is the Compton time. A quick way to see this is as follows: At the Compton scale we have,
$$|\vec {L} | = | \vec {r} \times \vec {p} | = | \frac{\hbar}{2mc} \cdot mc| = \frac{\hbar}{2},$$
that is, we get the Quantum Mechanical spin. Next, we can easily verify, that the choice,
\begin{equation}
t = \left(\begin{array}{ll}
1 \quad 0\\
0 \quad -1
\end{array}
\right), \vec {x} = \left(\begin{array}{ll}
0 \quad \vec {\sigma}\\
\vec {\sigma} \quad 0
\end{array}
\right)\label{ed}
\end{equation}
provides a representation for the coordinates in (3), apart from scalar factors. As can be seen, this is also a representation of the Dirac matrices. Substitution of the above in (\ref{e9a}) leads to the Dirac equation
$$(\gamma^\mu p_\mu - mc^2) \psi = 0$$
because
$$E\psi = \frac{1}{\epsilon}\{\psi' (x_j) - \psi (x_j)\}, \quad E = mc^2,$$
where $\epsilon = \tau$ (Cf.ref.\cite{wolf}).\\
Indeed, as noted, Dirac himself had
realized that his electron equation needed an average over spacetime
intervals of the order of the Compton scale to remove
zitterbewegung effects and give meaningful physics. This again is symptomatic of an underlying fuzzy
spacetime described by a noncommutative space time geometry
(\ref{eA}) or (4) \cite{sakharov}.\\
The point here is that under equation (\ref{eA}) and (\ref{ed}), the coordinates
$x^\mu \to \gamma^{(\mu)} x^{(\mu)}$ where the brackets with the
superscript denote the fact that there is no summation over the
indices.  Infact, in the theory of the Dirac equation it is well
known \cite{bade}that,
\begin{equation}
\gamma^k \gamma^l + \gamma^l \gamma^k = - 2g^{kl}I\label{e10a}
\end{equation}
where $\gamma$'s satisfy the usual Clifford algebra of the Dirac
matrices, and can be represented by
\begin{equation}
\gamma^k = \sqrt{2} \left(\begin{array}{ll}
0 \quad \sigma^k \\
\sigma^{k*} \quad 0
\end{array}\right)\label{e11a}
\end{equation}
where $\sigma$'s are the Pauli matrices. As noted by Bade and
Jehle (Cf.ref.\cite{bade}), we could take the $\sigma$'s or
$\gamma$'s in (\ref{e11a}) and (\ref{e10a}) as the components of a
contravariant world vector, or equivalently we could take them to
be fixed matrices, and to maintain covariance, to attribute new
transformation properties to the wave function, which now becomes
a spinor (or bi-spinor). This latter has been the traditional
route, because of which the Dirac wave function has its
bi-spinorial character. In this latter case, the coordinates
retain their usual commutative or point character. It is only when we
consider the equivalent former alternative, that we return to the
noncommutative
geometry (\ref{eA}).\\
That is, in the usual commutative spacetime the Dirac spinorial
wave functions conceal the noncommutative
character (\ref{eA}).
\section{Discussion}
We make a few brief comments. As noted earlier if terms of the order of $l^2$ are neglected in the above discussion, we return to ordinary non relativistic Quantum Mechanics. From Snyder's relativistically covariant relations (\ref{e3z}) we can see that order of $l^2$ terms need to be retained for relativistic theory, and it is no surprise that the Dirac equation then can be deduced. What we are seeling here is that for a relativistic theory there is a quantum of area of the order $l^2$ that is fundamental (Cf. also ref.\cite{baez}).\\
We have also noted that the usual angular momentum at the Compton scale which we get for example in the noncommutative relations of (\ref{e3z}) when order $l^2$ terms are retained, represents spin. Another way of looking at this is that from the Dirac theory we have (Cf. \cite{pdirac} and \cite{sch})
$$\vec{x} = (c^2 \vec{p} / H) t + \hat{x}$$
\begin{equation}
c\vec \alpha = \frac{c^2\vec p}{H} - \frac{2\imath}{\hbar} \hat x H\label{e12a}
\end{equation}
In (\ref{e12a}), the first terms on the right give the usual position and momentum, while the second term is the ``extra coordinate'' due to zitterbewegung. 
The angular momentum at the Compton scale is now given by $\sim \hat{x} \times \vec{p}$
\begin{equation}
(\hat{x} \times \vec{p})_z = \frac{c}{E} (\vec \alpha \times \vec{p})_z = \frac{c}{E} (p_2\alpha_1 - p_1\alpha_2)\label{ec}
\end{equation}
and similar equation for the other components. In (\ref{ec}) $E$ is the eigen value of the Hamiltonian operator. It shows that the usual angular momentum, but in the context of the Compton scale leads to the mysterious Quantum Mechanical spin contained in the $\vec{\alpha}$ matrices (Cf.ref.\cite{sakharov}).\\
Finally we remark that given the noncommutative geometry (\ref{eA}), it is possible to obtain a unified description of interactions \cite{uof}. 

\end{document}